\documentclass[12pt]{iopart}
\usepackage{graphicx}

\begin{document}
\title{Spectrally resolved quantum tomography of polarization entangled states}

\author{Dmitry A. Kalashnikov and Leonid A. Krivitsky}

\address{Data Storage Institute, Agency for Science, Technology and Research (A*STAR), 5 Engineering Drive I, 117608 Singapore}

\ead{Leonid\_Krivitskiy@dsi.a-star.edu.sg}

\begin{abstract}
{We study broadband polarization entangled states generated within the line width of type-II Spontaneous Parametric Down Conversion (SPDC). Applying a complete quantum polarization tomography protocol at arbitrary combination of frequency and angular sideband modes we reveal a complex structure of non-degenerate non-collinear polarization entangled states generated within the SPDC line width. It is demonstrated that simultaneous compensation of longitudinal and transverse walk off leads to homogenizing the structure of polarization entanglement.}
\end{abstract}
\maketitle 

\section{Introduction}
The concept of quantum entanglement which has been considered as a subject of purely theoretical interest some decades ago is nowadays accessible in many experimental labs ~\cite{nielsen}. Exploiting the process of Spontaneous Parametric Down Conversion (SPDC) allows obtaining a broad class of photon entangled states in both continuous and discrete degrees of freedom ~\cite{klyshko}. The feasibility of such experiments has opened a range of promising applications spanning from fundamental tasks such as tests of the concepts of quantum mechanics ~\cite{genovese} to nearly practical applications such as quantum cryptography ~\cite{gisin}, absolute metrology ~\cite{migdall}, quantum imaging ~\cite{lugiato} and quantum computation ~\cite{kok}.

 Within a broad class of entangled states generated via SPDC, polarization entangled states are of particular interest. Up to date, there are many developed and readily accessible methods to generate such states. The most straightforward method relies on type-II frequency degenerate SPDC in the collinear or non-collinear regime, when a photon of a laser pump decays into a pair of orthogonally polarized photons which are superimposed in well defined spatial modes ~\cite{shih,rubin,kwiat,kurtsiefer}. Other widely implemented methods are based on interference of orthogonally polarized downconverted pairs produced in two type-I SPDC crystals ~\cite{kwiat1,kim,burlakov, kwiat2} and a double pass of the pump through a single type-I crystal ~\cite{howell,dariano}.

In the present paper we consider generation of polarization entangled states in frequency degenerate collinear type-II SPDC from the continuous wave pump ~\cite{shih,rubin}. Due to the limited length of SPDC crystal, the phase matching condition imposes a spread of correlated frequency and angular modes in the vicinity of exact degeneracy, further referred to as \textsl{SPDC line width}. In this case the polarization entangled two-photon state that occupies correlated sideband modes A and B, further referred to as \textsl{sidebands}, can be written as follows: $\left|\Psi\right\rangle\propto\left|H_{A}V_{B}\right\rangle+e^{i\phi}\left|V_{A}H_{B}\right\rangle$, where indices H and V denote horizontal and vertical polarizations of photons, respectively and $\phi$ is a relative phase. Considering the birefringence and chromatic dispersion of SPDC crystal, it can be shown that polarization states at different angular and frequency sidebands acquire different relative phase $\phi$ ~\cite{brida}. This effect, also referred to as longitudinal (for frequency modes) and transverse (for angular modes) walk-off, leads to generation of  inhomogeneous broadband entangled state within the SPDC line width, which prevents observation of high-visibility two-photon polarization interference. Several techniques have been suggested to restore the two-photon interference in type-II SPDC. Some are based on filtering  SPDC spectrum, however significantly reducing the brightness of the source ~\cite{rubin}. Others are based on compensation of the walk-off by means of a specifically tailored birefringent material placed into a downconversion beam ~\cite{rubin,kwiat}.

Despite the fact that the majority of experiments aim to eliminate the walk-off in order to optimize generation of a given entangled state, it is highly intriguing that in the case when the walk-off is not compensated, a single nonlinear crystal can be used as a natural source of different entangled states. In particular, it allows simultaneous generation of slightly mismatched, however, completely orthogonal entangled states. Note also, that the above effect should always be accurately considered in observation of two-photon interference with narrow band filters as it would define the purity and entanglement of post-selected states.

A structure of polarization entangled states generated within the SPDC line width has been considered earlier in ~\cite{brida}. However, the experiments have been performed separately for frequency degenerate and strictly collinear regimes and did not allow accessing any non-degenerate non-collinear states. Apart from that, characterization of the states was limited only to observation of basic polarization dependencies which, as it will be shown later, do not allow a complete state characterization. In the present work, we make an effort to fill these gaps and present a complex approach, where by using quantum polarization tomography we are able to reconstruct a density matrix of the entangled state in an arbitrary combination of angular and frequency SPDC sideband modes.

\section{Theory}

Let us consider type-II SPDC in a critically phase matched birefringent crystal of length $L$, pumped by a continuous wave laser with frequency $\Omega_{p}$  and wave vector $\vec{k_{p}}$. The orientation of the optical axis is set in such a way that orthogonally polarized signal and idler photons $(s,i)$ have nearly degenerate frequencies ($\Omega_{s}\approx\Omega_{i}\approx\Omega_{p}/2$) and propagate collinearly ($\vec{k_{s}}||\vec{k_{i}}||\vec{k_{p}}$), yielding energy and momentum conservation:

\begin{eqnarray}
\Omega_{s}+\Omega_{i}=\Omega_{p};  
\vec{k_{s}}+\vec{k_{i}}\approx\vec{k_{p}};\nonumber
\end{eqnarray}
Further it is assumed that the exact momentum conservation is fulfilled in the transverse direction of the crystal (orthogonal to $\vec{k_{p}}$), provided that the pump diameter is larger than the transverse walk-off, given by $L\tan\theta$, where $\theta$ is a typical scattering angle. In this case, the finite length of the SPDC crystal in the longitudinal direction (along $\vec{k_{p}}$) weakens the momentum conservation condition and allows generation of correlated photons in non-degenerate sideband modes. Expanding the longitudinal component of the momentum mismatch $\Delta_{z}$  up to linear terms in frequency ($\omega=\Omega_{s}-\Omega_{p}/2$) and angular ($\theta$) sidebands, one obtains ~\cite{thecase,burlakov1}:

\begin{eqnarray}
	\vec{\Delta}\equiv{\vec{k_{p}}-\vec{k_{s}}-\vec{k_{i}}}; \Delta_{z}\approx D\omega + B\theta,\\
	D=(dk_{e}/{d\omega}-dk_{o}/{d\omega}); B=dk_{e}/{d\theta}  \nonumber
\end{eqnarray}
where $k_{o,e}$ is the wave vector for ordinary (horizontally polarized) and extraordinary (vertically polarized) waves in the crystal, respectively. In this case, the two photon polarization entangled state, generated within the whole SPDC line width, can be written as follows:

\begin{equation}
	\left|\Psi\right\rangle\propto\int\int d\omega d\theta F(\omega,\theta)\left(\left|H^{\theta}_{\omega}V^{-\theta}_{-\omega}\right\rangle+e^{i\phi}\left|V^{\theta}_{\omega}H^{-\theta}_{-\omega}\right\rangle\right)
\end{equation}
where $|H^{\theta}_{\omega}V^{-\theta}_{-\omega}>$ ($|V^{\theta}_{\omega}H^{-\theta}_{-\omega}>$) is a two-photon Fock state with one horizontally (vertically) polarized  photon in mode ($\omega;\theta$) and one vertically (horizontally) polarized  photon in mode ($-\omega;-\theta$) and $\phi\equiv\Delta_{z} L$ is a relative phase. The function $F(\omega,\theta)$  represents a  spectral amplitude of type-II SPDC process, which is given by ~\cite{rubin}:

\begin{equation}
	F(\omega,\theta)=\sin(\Delta_{z} L/2)/(\Delta_{z} L/2)\equiv\rm{sinc}(\Delta_{z} L/2).
\end{equation} 
 
For the future discussion it is of crucial importance to consider a dependence of the relative phase $\phi$ in (2) on  frequency and angular sidebands. In Fig.1a we present calculations of SPDC spectral amplitude (3) near the exact degeneracy  and  in Fig.1b a corresponding phase dependence, using (1). From Fig.1b, one can see that each combination of correlated frequency and angular sidebands $(\omega;\theta)$ defines a specific value of the relative phase $\phi$ in (2) provided that a continuum of non-degenerate non-collinear entangled states is generated within the SPDC line width. Thus, for instance, even for non-degenerate modes, when a choice of specific sidebands results in $\phi=0$, one of the triplet $\Psi^{+}\equiv\left(\left|H^{\theta}_{\omega}V^{-\theta}_{-\omega}\right\rangle+\left|V^{\theta}_{\omega}H^{-\theta}_{-\omega}\right\rangle\right)/\sqrt{2}$  Bell states is generated. At the same time, when the combination of frequency and angular sidebands  results in $\phi=\pi$, then the singlet Bell state $\Psi^{-}\equiv\left(\left|H^{\theta}_{\omega}V^{-\theta}_{-\omega}\right\rangle-\left|V^{\theta}_{\omega}H^{-\theta}_{-\omega}\right\rangle\right)/\sqrt{2}$ is generated. 

Let us mention, that the above effect clearly explains the difficulty of observation of two-photon interference in case of a broadband detection. Indeed, from (1) and (2) it follows that a resultant state would represent a statistical mixture of pure states with different relative phases. In this case the entropy and entanglement of generated mixed state would depend on corresponding bandwidths of spatial and frequency filters, opening however the possibility to study mixed entangled states ~\cite{white}. In order to restore the two-photon interference, the phase dependence on scattering parameters should be erased. This can be done, for instance, by placing a piece of birefringent material after the SPDC crystal, which adds a phase shift $\phi_{comp}$ to the initial relative phase $\phi$. If the introduced phase shift flattens the dependence of the total phase $\phi_{total}\equiv\phi+\phi_{comp}$ on both scattering parameters within the detection band, then the two-photon interference can be clearly observed.

Let us now consider the general representation of a  two-photon polarization state. The following  set of four orthogonal basis states can be defined for two distinguishable photons in the horizontal-vertical polarization basis: $\left|H^{\theta}_{\omega}H^{-\theta}_{-\omega}\right\rangle;\left|H^{\theta}_{\omega}V^{-\theta}_{-\omega}\right\rangle;\left|V^{\theta}_{\omega}H^{-\theta}_{-\omega}\right\rangle;\left|V^{\theta}_{\omega}V^{-\theta}_{-\omega}\right\rangle$ ~\cite{bogdanov}. Therefore, an arbitrary  two-photon polarization state in given sidebands can be written as a state of a four level quantum system (ququart) of the following form:
\begin{equation}
\left|\Psi\right\rangle=c_{1}\left|H^{\theta}_{\omega}H^{-\theta}_{-\omega}\right\rangle+c_{2}\left|H^{\theta}_{\omega}V^{-\theta}_{-\omega}\right\rangle+c_{3}\left|V^{\theta}_{\omega}H^{-\theta}_{-\omega}\right\rangle+c_{4}\left|V^{\theta}_{\omega}V^{-\theta}_{-\omega}\right\rangle,
\end{equation}
where $c_{i}=|c_{i}|\exp(i\phi_{i})$ is the complex probability amplitude of the corresponding two-photon state, fulfilling the normalization condition $\sum{\left|c_{i}\right|}^2=1$. In the ququart basis the state (2) has the following complex amplitudes: $c_{1}=0;c_{2}=1/\sqrt{2};c_{3}=\exp({i\phi})/\sqrt{2};c_{4}=0$. Note, that (2) can be easily transformed into an arbitrary state (4) by polarization rotations on separate photons, and in particular into the triplet Bell states $\Phi^{\pm}\equiv\left(\left|H^{\theta}_{\omega}H^{-\theta}_{-\omega}\right\rangle\pm\left|V^{\theta}_{\omega}V^{-\theta}_{-\omega}\right\rangle\right)/\sqrt{2}$ ~\cite{kwiat}. 
Generation, transformation, and measurement of ququarts have been extensively studied in earlier works in strongly non-frequency degenerate ~\cite{bogdanov,trojek} and in non-collinear regimes ~\cite{james,poh}. At the same time, exploiting polarization entanglement within the SPDC line width allows for the first time, to the best of our knowledge, obtaining ququart states in non-frequency degenerate, non-collinear modes.

\begin{figure}
\begin{center}
\includegraphics[width=6cm,height=6cm]{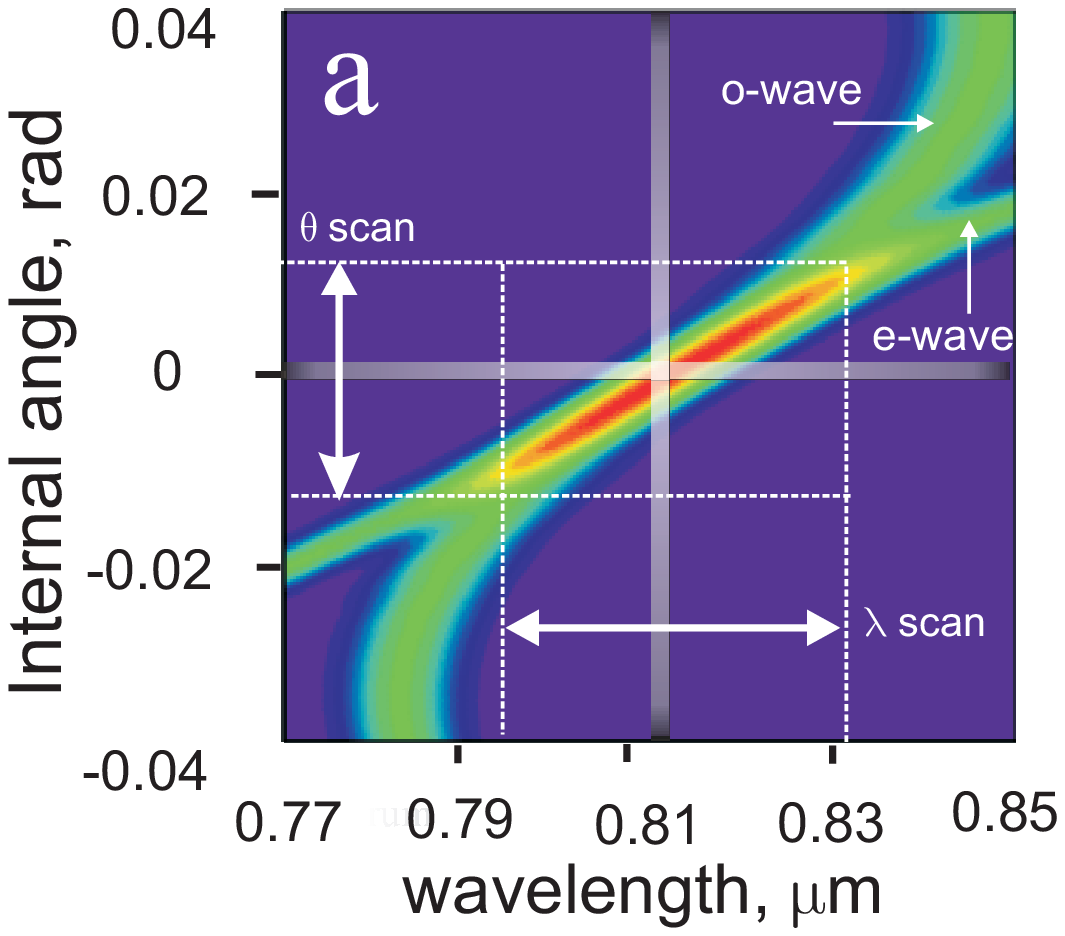}
\includegraphics[width=8cm,height=8cm]{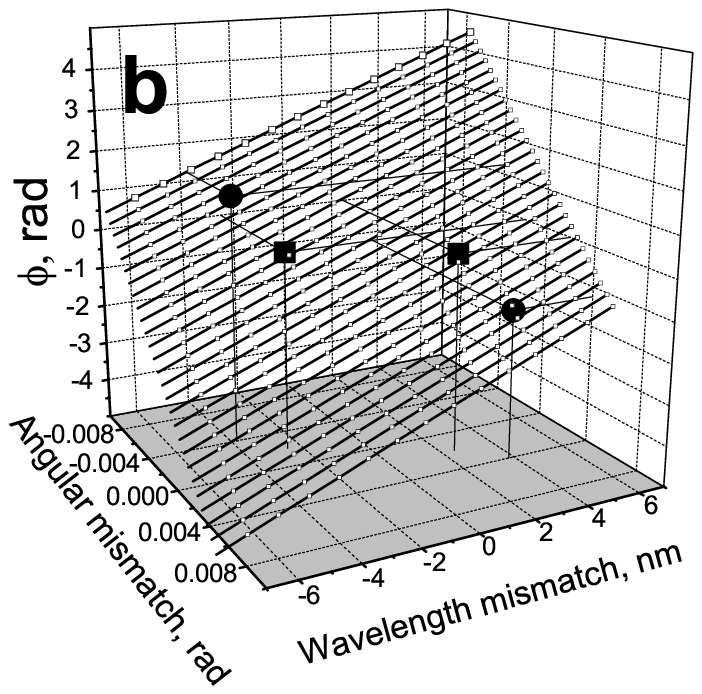}
\caption{(Color on-line)(a) Calculation of the spectral distribution according to (3) with the parameters corresponding to the described experiment in Section IV. The shaded stripes and arrows schematically show the detection bandwidths and tuning ranges of the detectors in angular (horizontal white shaded) and frequency (vertical white shaded) sidebands. (b) The dependence of the relative phase in (2) on the wavelength and angular mismatch. The pairs of black squares and black circles depict non-degenerate correlated modes which were chosen for tomography reconstruction in Section V.}
\end{center}
\end{figure}

\section{Measurement Procedure}

Polarization entangled state (2) can be experimentally characterized with a simple setup that consists of a 50/50 non-polarizing beam splitter, two polarization analyzers installed at its output ports and two photodetectors ~\cite{kwiat}. Two complimentary measurements of coincidences in configuration when one of the analyzers is oriented at 45 degrees to laboratory basis and another analyzer is flipped between 45 and -45 degrees, would allow to measure the relative phase $\phi$, according to the formulas~\cite{brida}:
\begin{eqnarray}
	R^{45/45}_{c}=\rm{sinc}^{2}(\phi/2)\cos^{2}(\phi/2) \\ \nonumber
	R^{45/-45}_{c}=\rm{sinc}^{2}(\phi/2)\sin^{2}(\phi/2).
\end{eqnarray}
Despite the fact that the above measurements provide a comprehensive and rapid estimation of the state, they do not allow complete state reconstruction. When one considers an entangled state (2) as a particular case of a ququart state (4), a more sophisticated procedure is needed that would allow reconstruction of 16 density matrix elements in (4). 

In the present work different entangled states at arbitrary combinations of angular and frequency sidebands are characterized by means of a complete protocol of quantum polarization tomography. The quantum polarization tomography is an indispensable tool for characterization of polarization entangled states, allowing complete reconstruction of the density matrix via correlation measurements in various polarization bases ~\cite{james,bogdanov1,rehacek}. Among the variety of suggested protocols we exploit a particular one by James et al.~\cite{james}, which is referred to as J16 ~\cite{deburgh}. The choice of J16 is motivated by the convenience of its experimental implementation to study slightly mismatched entangled states. Thus, in contrast to the method by Bogdanov et al. ~\cite{bogdanov1}, it does not require simultaneous wavelength dependent transformations on both photons. According to the chosen protocol, a wide aperture broadband 50/50 non-polarizing beamsplitter is placed into the downconverted beam with two sets of polarization analyzers in its output ports, followed by single-photon detectors. Each polarization analyzer consists of broadband quarter- and half- wave plate and a linear polarization filter, which allows projection of a polarization state of a single photon onto an arbitrary state on a Poincare sphere. Coincidence measurements between two detectors at 16 pre defined orientations of the wave plates allow a complete reconstruction of the density matrix of unknown state (4).

To allow an arbitrary choice of frequency-angular mode within the SPDC line width, a traditional tomography set-up has been upgraded in the following way. One detector selects a narrow band in a frequency spectrum, and at the same time, accepts a whole angular spectrum. Similarly, another detector performs a narrow angular selection, while the frequency spectrum is not filtered. In Fig.1a we schematically depict the corresponding bandwidths and tuning ranges of two detectors by horizontal white shaded (angular selective detector) and vertical white shaded (frequency selective detector) stripes. For the case of a not strongly focused continuous wave pump such configuration allows filtering of an arbitrary polarization state by tuning a frequency filter in one arm and a spatial filter in another arm. Note, that a broadband detection of a conjugated mode allows to avoid a drastic decrease in the signal and thus reduces a contribution of dark counts of the detectors.

\section{Experiment}

The experimental setup on polarization tomography of SPDC sidebands is shown in Fig.2. A pump beam from a cw diode laser with the power 50mW at the wavelength of 406 nm and a spectral full width half maximum (FWHM) 0.3 nm, was mode cleaned by a single mode fiber and was focused by a lens (\textit{f}= 500 mm) into a 1mm long BBO crystal cut for type-II SPDC. The waist of the pump beam in the crystal was 65 $\mu m$, large enough to have no significant effect on broadening the SPDC spectrum. The crystal was set for collinear frequency degenerate regime of SPDC with its axis at 47.6 degrees to $\vec{k_{p}}$, so that both signal and idler photons had the same central wavelength of 812 nm and propagated collinearly. After passing through the crystal, the pump was rejected by a UV-mirror (UVM) while SPDC radiation fluently passed through it and then was divided into two paths by a 50/50 non-polarizing beam splitter (NPBS). In each output port of NPBS a set of broadband quarter- and half- wave plates (HWP) projected the state onto a pre-defined linear polarization state (the horizontal one), selected by a polarizing beam splitter (PBS). In one output arm of NPBS the whole SPDC angular line width, estimated as 0.01 rad FWHM, see Fig.1, was imaged with a demagnification (M=12) at the face of a single-mode fiber (SMF) with a numerical aperture 0.12. The SMF led into a grating spectrometer with a resolution of 0.35 nm. In another port of NPBS, a lens L1 (\textit{f}=500 mm)  was placed in such a way that SPDC crystal was at its front focus. Thus, the lens performed a projection of scattering angles into physical space, where they were filtered with a 1 mm diameter translatable aperture (A). This scheme provided an angular resolution of $1\cdot10^{-3}$ rad. A broadband interference filter (FWHM=100nm) was used to pass the whole spectral band of SPDC whilst cutting off the undesirable optical noise. The Perkin-Elmer SPCM-14FC avalanche photodiodes (D1, D2) were used to detect photons in each port. Signals from both photodiodes were then sent to a coincidence unit (\&) with a time window of 5 ns.

For the reconstruction of density matrix we used the above mentioned tomography protocol J16 ~\cite{james}, where 16 sets of orientations of wave-plates provided projections at a-priori defined polarization states. At each set 20  measurements were performed with the acquisition time of 10 seconds. The number of coincidences obtained at each setting corresponded to the linear combination of components of the density matrix. Thus after the tomography procedure one obtains a system of 16 equations for 16 unknown parameters ~\cite{tool}.

\begin{figure}
\begin{center}
\includegraphics[width=12cm]{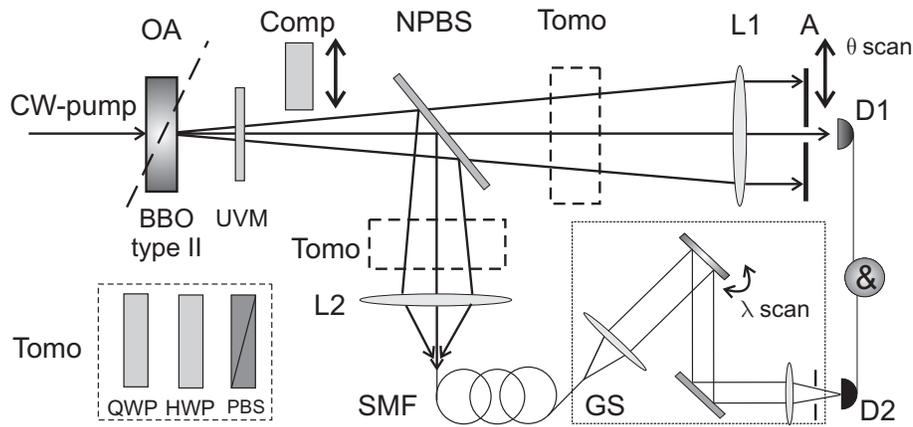}
\caption{Experimental setup: a 406nm cw laser is focused into 1 mm long type II BBO and filtered out by UV mirror (UVM). Broadband SPDC radiation is split by a 50/50 non-polarizing beam splitter (NPBS) and sent into two tomography analyzers (TOMO), consisting of a quarter- and half- wave plates (QWP, HWP) and a polarizing beam splitter (PBS). One branch of NPBS is used for angular scan and has a 500mm lens (L1) focused at the crystal and a 1mm aperture (A), mounted onto a translation stage.  Another branch of the NPBS accepts a broad angular spectrum in a single mode (SM) fiber, which is connected to a grating spectrometer (GS). D1, D2 are avalanche photodiodes, connected to  coincidence scheme (\&). A quartz compensator (Comp) was used in some experiments to demonstrate homogenization of polarization entanglement.}
\end{center}
\end{figure}

In experiments devoted to study the walk-off compensation, a specially prepared compensator was introduced just after the UVM. The parameters of the compensator, i.e. length and orientation of its optical axis were chosen in such a way that the introduced phase $\phi_{comp}$ being added to the original phase $\phi$ at any combination of angular and frequency sideband resulted in $\phi_{total}=\phi+\phi_{comp}\approx0$, thus providing a homogeneous preparation of the quantum state. A 6.5 mm long piece of crystalline quartz was used for this purpose with its optical axis orientated at 49.6 deg with respect to the wave vector of the pump.

\section{Results and discussion}

First, we have checked that the set-up allows broadband registration and has a good enough spectral resolution to study entangled states generated within the frequency and angular SPDC line width. Angular and frequency scans were performed in natural (filtering H/V polarizations by the tomography block) and diagonal polarization bases (filtering 45/-45 deg polarizations) in frequency degenerate  and collinear cases, respectively. The interference pattern presented in Fig.3 clearly demonstrates the inhomogeneous structure of polarization entanglement in angular (a,b) and frequency (c,d) sidebands. This confirms that a broad range of states is generated with a relative phase determined by both frequency and angular offset from the exact degeneracy. By introducing the quartz compensator, the relative phase in (2) changes to $\phi_{total}=\phi+\phi_{comp}\approx0$ for any side band. Thus, a simultaneous homogenization of the interference pattern in both angular and frequency spectra was observed, providing generation of the Bell $\Psi^{+}$ state within the whole spectrum. 

\begin{figure}
\includegraphics[width=8cm]{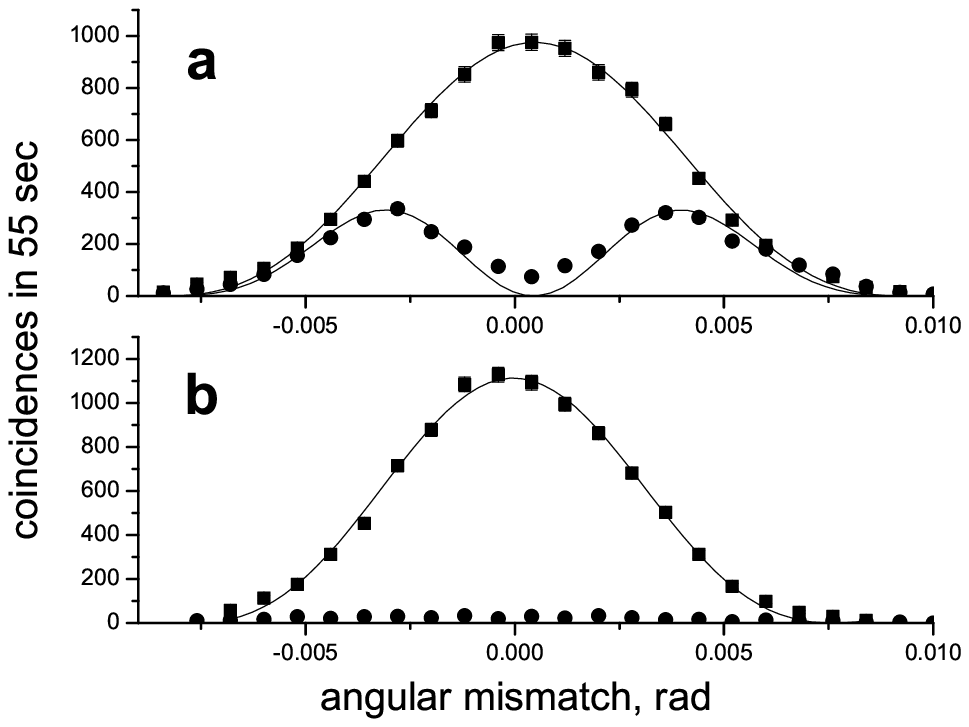}
\includegraphics[width=8cm]{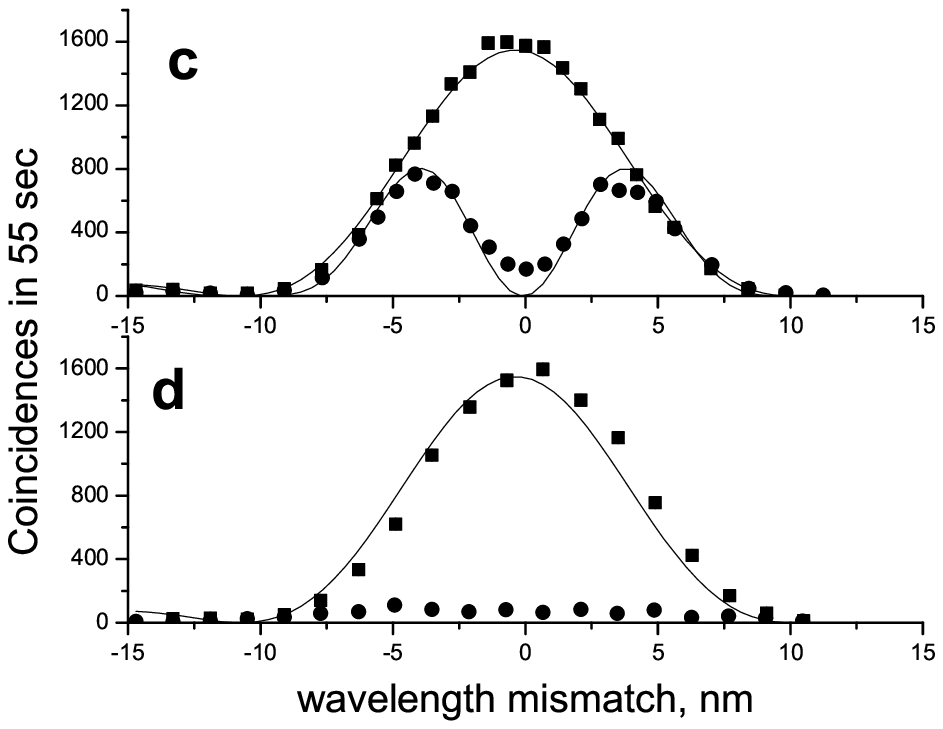}

\caption{Angular (a) and frequency (c) spectra without the compensator in natural H/V (squares) and diagonal 45/-45 basis (circles). The inhomogeneous structure of polarization entanglement is clearly seen and is due to variation in phase in (2). Compensation for longitudinal and transverse walk-off by means of a specially designed compensator allows homogenization of polarization entangled states in angular (b) and frequency (d) spectra, so that $\Psi^{+}$ state is generated within the whole SPDC line width. The solid curves are fitted according to (3) (squares) and (5) (circles) }

\end{figure}

In order to stress that the complex polarization analysis can be performed on the set-up, one can choose the state to reconstruct in such a way that it should not be at trivial settings such as a collinear or frequency-degenerate regime. First, the state is considered at a pair of conjugated wavelengths 814.5 nm and 809.5 nm and at angular modes $\theta=\pm$ 0.002 rad. According to the theoretical predictions in Fig.1b, where the chosen sidebands are shown as two squares, this arrangement corresponds to the triplet Bell $\Psi^{+}$ state to be filtered out of SPDC line width. Note that in this case, the observed signal is of the same level as in the degenerate mode, despite being detuned from the exact degeneracy. The polarization tomography of the state has been performed at this particular frequency-angular sideband. After elaboration of the results, it was found that the raw reconstructed density matrix did not satisfy to the basic physics requirements, such as e.g. the positivity of its eigen values. A maximum likelihood estimation (MLE) algorithm was applied in order to find the maximally close physical state ~\cite{james,rehacek1, tool}. Thus, the following density matrix was obtained: 

\begin{equation}
\fl\rho_{0}=\left(\begin{array}{cccc}      
 0.0068& -0.0356-0.0127i& -0.0370+0.0004i& -0.0003+0.0005i\\
-0.0356+0.0127i&  0.4615& 0.4369-0.0258i& 0.0095+0.0225i\\
-0.0370-0.0004i& 0.4369+0.0258i& 0.5275& 0.0057+0.0243i\\
-0.0003-0.0005i&  0.0095-0.0225i& 0.0057-0.0243i& 0.0042
    \end{array}\right)
 \end{equation}
%
The experimental results on the reconstructed state are presented in Table 1 and the real and imaginary parts of the density matrix (6) are presented in Fig.4a,b. For all the experiments, the fidelity between the reconstructed state $\rho_{exp}$ and the theoretical state $\rho_{th}$ was calculated according to $F=(Tr(\sqrt{\sqrt{\rho_{th}}\rho_{exp}\sqrt{\rho_{th}}}))^2$.

The above protocol has been accomplished in this set-up, by placing a birefringent compensator after the UVM, which provides generation of the Bell $\Psi^{+}$ state, as well. The corresponding data on the reconstructed state showed the fidelity of experimental results $F=0.92\pm0.01$, which is in agreement with the results obtained without the compensator.  

Further more, states at an arbitrary combination of frequency and angular modes have been studied with the tomography setup. We have chosen the state at an angular mismatch of $\theta=\pm$ 0.004 rad, whilst a pair of correlated wavelengths has been changed to 815.8nm and 808.2 nm.  The chosen sidebands, which are shown in Fig.1b by a pair of black circles, correspond to the filtered state (2) with a phase $\phi=-6\pi/15$. The result of the state reconstruction after application of MLE is presented below, plotted in Fig.4c,d and analyzed in Table 1. 

\begin{equation}
\fl\rho_{\frac{-6\pi}{15}}=\left(\begin{array}{cccc}      
0.0073&            -0.0170 + 0.0146i&   0.0074 + 0.0108i&  -0.0028 - 0.0030i\\
  -0.0170 - 0.0146i&   0.4840&             0.1430 - 0.4071i&  -0.0010 - 0.0128i\\
   0.0074 - 0.0108i&   0.1430 + 0.4071i&   0.5043&             0.0181 - 0.0024i\\
  -0.0028 + 0.0030i&  -0.0010 + 0.0128i&   0.0181 + 0.0024i&   0.0044
      \end{array}\right)
  \end{equation}

\Table{Experimental results on tomography of polarization entangled states in different sidebands of SPDC line width. $\lambda_{s,i}$, are the selected wavelengths and $\theta_{s,i}$ are the internal scattering angles of signal and idler photons,respectively. The Fidelity is calculated between the experimentally reconstructed state and state (2) with a corresponding value of the relative phase $\phi$, defined by the selected sidebands.}
\br
Symbol&\centre{3}{Selected state}&\centre{2}{Tomography results}\\
\ns
in Fig.1b&\crule{3}&\crule{2}\\
&$\lambda_{s}; \lambda_{i},nm$ & $\theta_{s}; \theta_{i}$, rad & $\phi$, rad & Fidelity & $Tr(\rho)^{2}$\\
\mr
$\fullsquare$ & 809.5; 814.5 &  $\pm$0.002 & 0        & 0.931$\pm$0.015  & 0.8824 \\
$\fullcircle$ & 808.2; 815.8 &  $\pm$0.004 & -6$\pi$/15 & 0.926$\pm$0.015 & 0.8634 \\
\br
\end{tabular}
\end{indented}
\end{table}

With independent measurements (not shown), a slight difference was observed in spectral widths of orthogonally polarized photons for highly detuned angular sidebands. The observations from these experiments explain the inequality of diagonal components of the density matrix that leads to the degradation of the fidelity of the reconstructed state.

The analysis of experimental uncertainties of tomography procedure revealed that for the presented results, the value of fidelity and purity of the state was partially compromised by the final spectral resolution of angular and frequency scans and by the purity of the source itself. The finite resolution of the spectrometer and angular filter contributed to a non-homogeneity of a chosen phase, launched into the tomography setup. In its turn, the purity of the state, produced by the source, was affected by the final frequency bandwidth of the pump and diffraction of SPDC angular modes. Therefore, the chosen experimental parameters were optimized to find a reasonable compromise between the mentioned factors and a reasonable collection efficiency, to allow a reliable tomography reconstruction. 

\begin{figure}
\includegraphics[width=8cm]{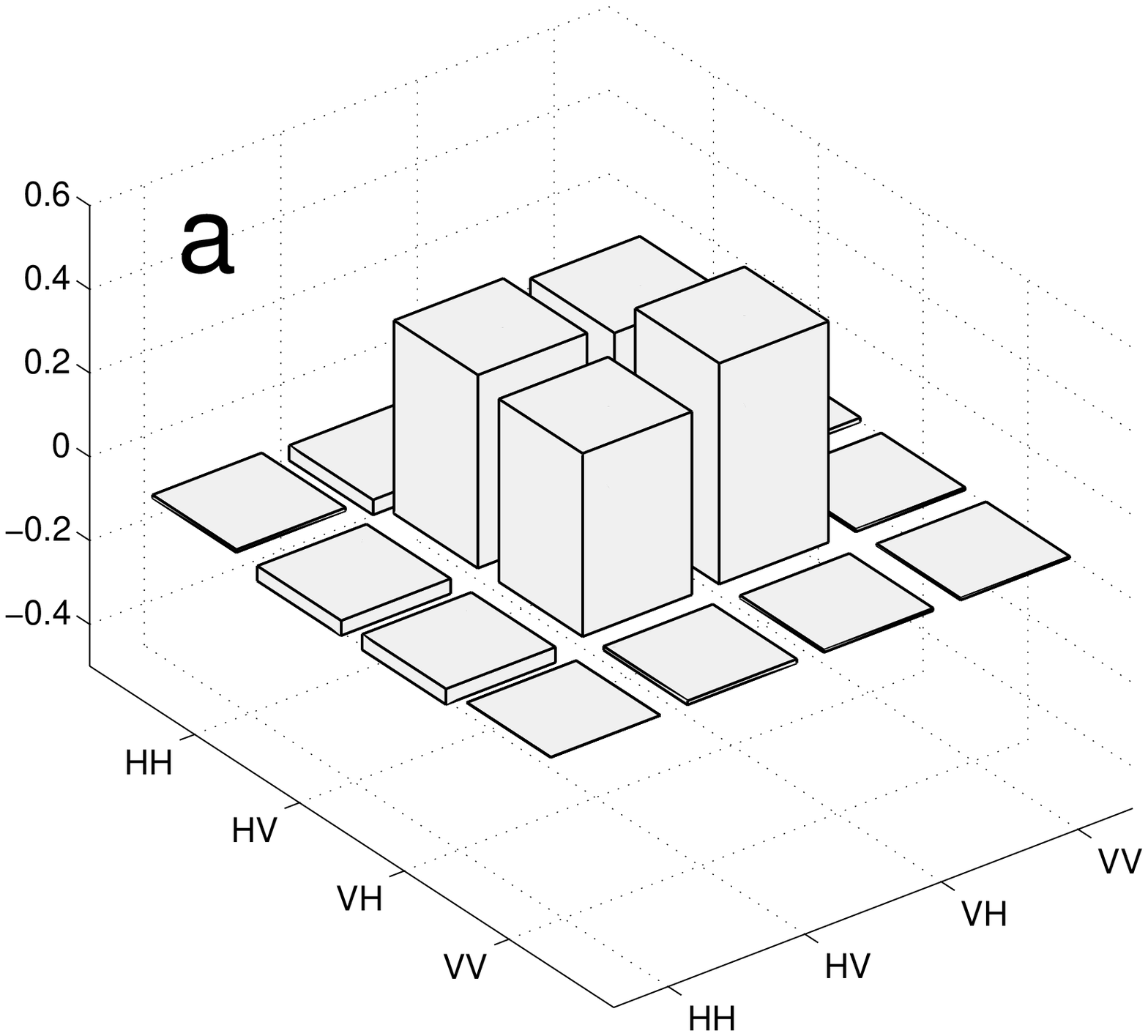}
\includegraphics[width=8cm]{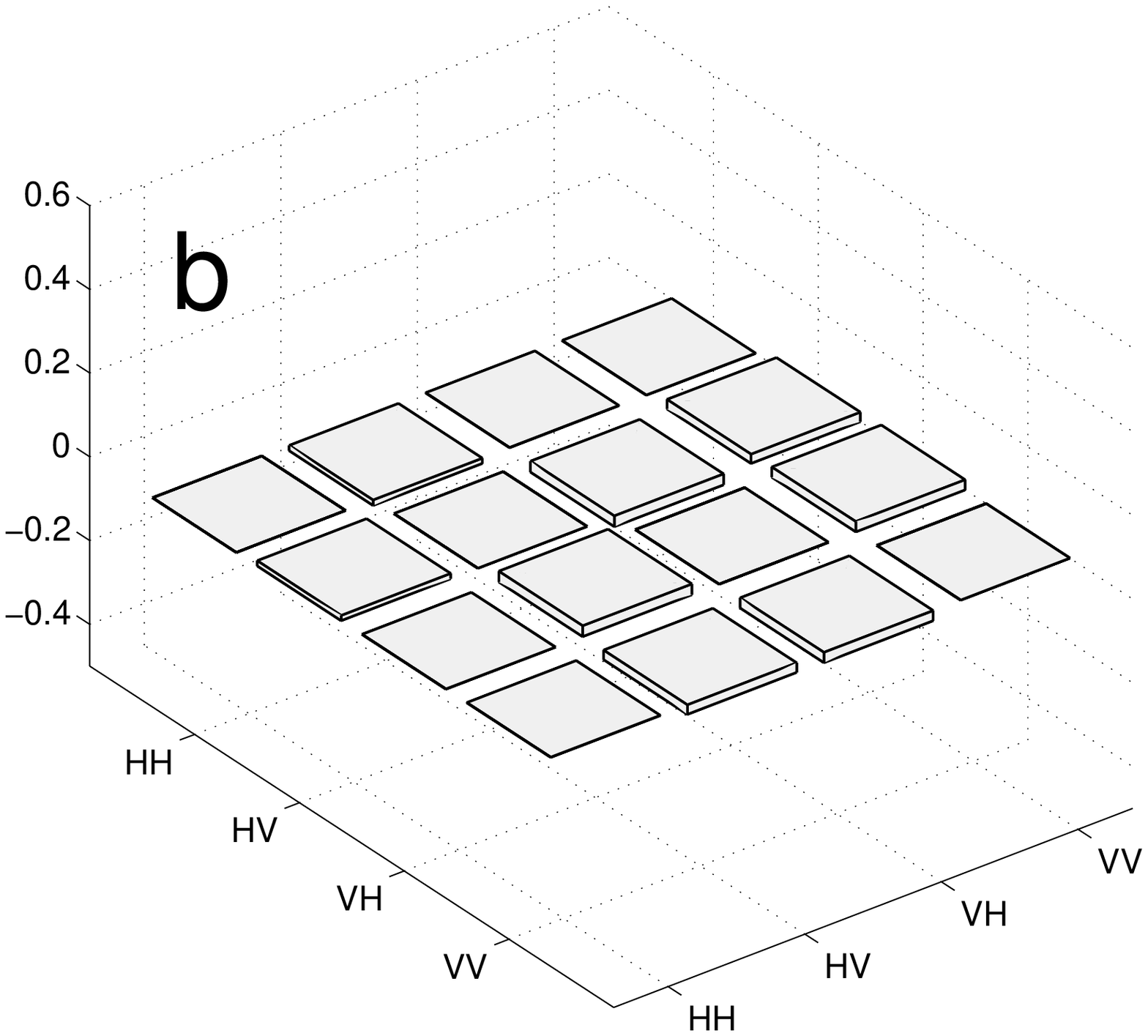}
\includegraphics[width=8cm]{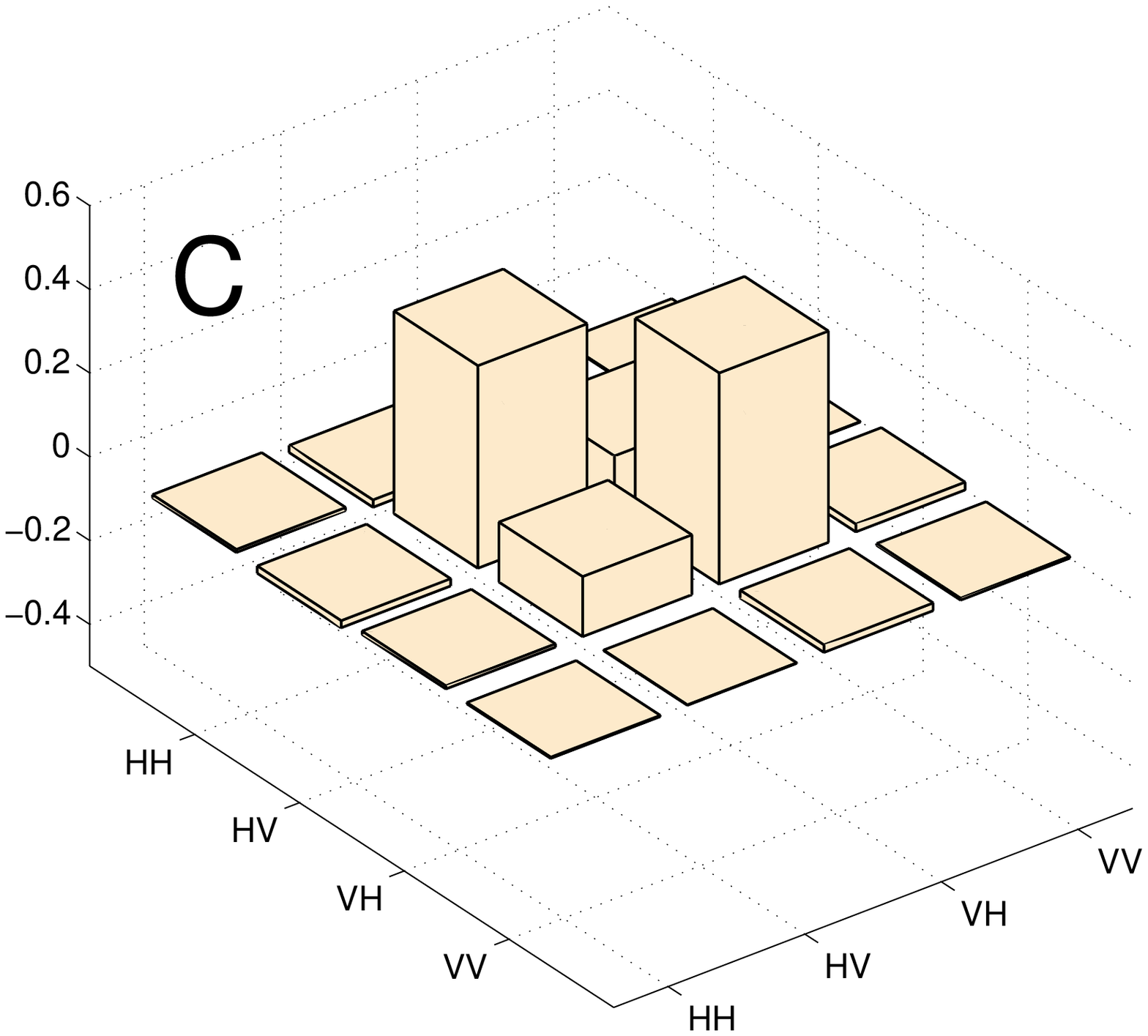}
\includegraphics[width=8cm]{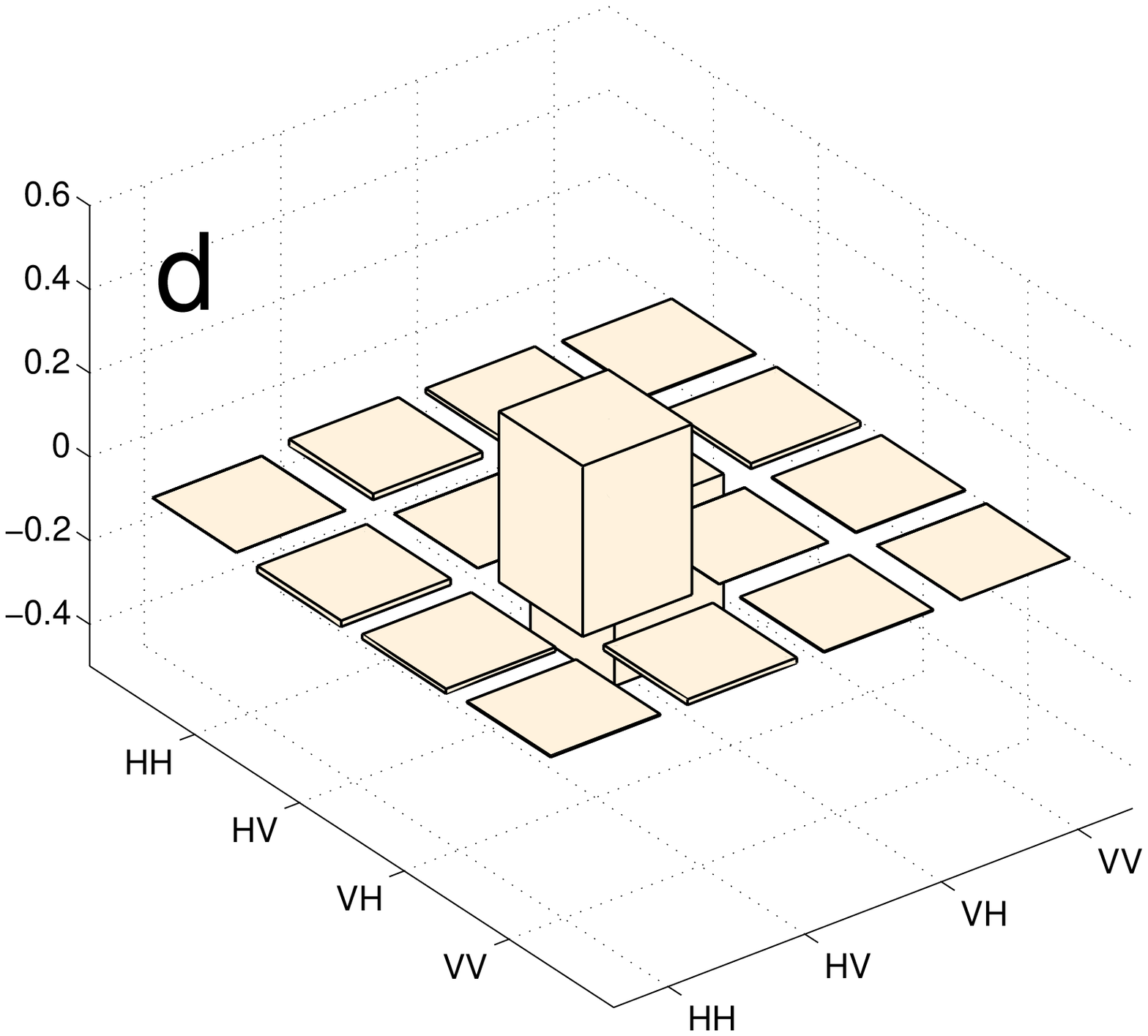}
\caption{Graphical representation of experimentally reconstructed density matrices. Fig.4a,b the real and imaginary parts of (6), respectively; Fig4.c,d the real and imaginary parts of (7), respectively.}

\end{figure}

\section{Conclusions}
In conclusion, the variety of non-degenerate non-collinear polarization-entangled states, generated within the natural line width of type-II SPDC has been studied with a quantum tomography setup. It has been demonstrated that due to a multi-mode nature of SPDC process the structure of polarization entanglement is inhomogeneous. Our results suggest that in experiments with type-II SPDC an accurate analysis of down converted modes and their spectral bandwidths is essential for ensuring high purity and entanglement of produced states. In addition, it has been demonstrated that  inhomogeneity of polarization entanglement can be effectively tailored with a birefringent compensator. In this regard, it is of practical interest when polarization entanglement is simultaneously homogenized in spatial and frequency domains, providing generation of a given broadband entangled state at high photon fluxes. It is also worth mentioning that the developed technique will allow obtaining a limited class of mixed states with their entropy being controlled by the settings of the experiment (e.g. by filter bandwidth). This opportunity represents a considerable interest for future experimental studies of mixed entangled  states ~\cite{white}.

\section{Acknowledgments}
We acknowledge M. Chekhova for helpful discussions and G. Maslennikov for advice on experimental details. This work was supported by the A*Investigatorship grant.


\begin{thebibliography}{<30>}
\bibitem{thecase} Here we consider the scattering in the plane, containing the optical axis of the crystal. In case of the scattering in the isotropic plane of the crystal a uniform entangled state is generated within the line width. See Ref. [18] for more details.
\end{thebibliography}
\end{document}